# Image Classification Algorithm for Determining the Light Curve Morphologies of *ASAS-SN* Eclipsing Binaries

Burak Ulaş[1, 2, a)]

[1]*Department of Space Sciences and Technologies, Faculty of Arts and Sciences, Çanakkale Onsekiz Mart University, Terzioğlu Campus, TR-17100, Çanakkale, Turkey*
[2]*Çanakkale Onsekiz Mart University, Astrophysics Research Center and Ulupınar Observatory, TR-17100, Çanakkale, Turkey*

[a)]Corresponding author: burak.ulas@comu.edu.tr

**Abstract.** We present a classification of the light curve morphologies of eclipsing binary systems observed by ASAS-SN based on their light curve images. The data of 16500 eclipsing systems having three different classes (detached Algol-type, β Lyr type, and W UMa type) are collected to construct their light curves. A deep learning algorithm containing the convolutional neural networks is employed on the images to achieve a satisfying classification. A code called ASEBCLASS is written in Python language and it uses the TensorFlow platform through Keras API to employ the mathematical libraries in the training phase of the model. The architecture consists of four groups of convolutional, activation, maximum pooling layers together with the additional fully connected layers. The results show that our algorithm estimates the morphological class of an external input image data with an accuracy value of 92%.

## INTRODUCTION

Machine learning algorithms gained ground in various scientific applications, especially in the last few decades. As a subclass of machine learning, the deep learning algorithms made the scientists be able to construct neural network models giving relatively more accurate results in a shorter time duration in their scientific investigations dealing with a large amount of data when comparing to the traditional, even the conventional machine learning methods [1]. Since the performance of the deep learning algorithms exceeded the manmade applications [2], they expanded in a very large variety of research fields carrying out the huge datasets. Neural networks in deep learning algorithms try to simulate the simple model of biological neuron structure and learn from the training data. The image classification algorithm, which is used in this study, is a good example among the applications using the neural networks, more precisely the convolutional neural networks [3]. These types of networks have their origins from the studies like [4] in the late eighties.

In an image classification deep learning algorithm, the layer transforms the input data based on its parameters. Then predictions calculated from the output of the layer are compared to the expected output and a loss score is calculated. The algorithm uses an optimizer to adjust weights, start over the transformation, and lower the loss score [5]. A brief representation of the process is shown in Fig. 1.

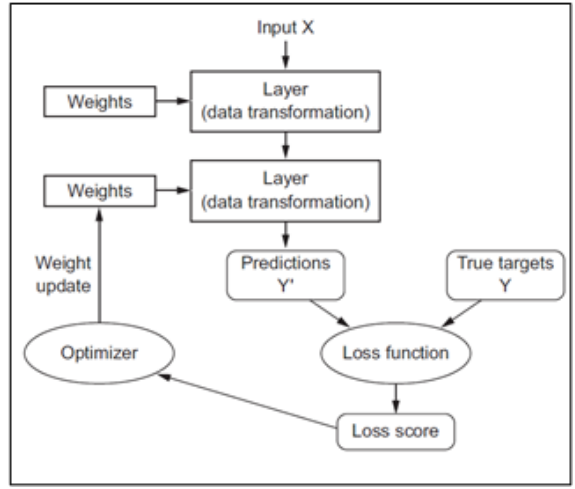

**FIGURE 1.** Main steps of image classification using deep learning algorithm as given by [5].

ASAS-SN (All Sky Automated Survey for Supernovae) [6-9] is a survey project that focused on supernova phenomena; therefore, numerous supernovae were discovered by the project. Since a big part of the sky is observed during the survey various types of stars also fall in its target. The observational data of the project were released on the project's web page. The project consists of data from the observations of 24 telescopes around the world. These instruments can achieve $18^{th}$ magnitude which is 50000 times greater than the human eye. The variable star section of the page presents the properties of 666502 variable stars as well as their light curve data. The full dataset is also downloadable through the webpage.

This study contains three main sections: light curve image production from collected data, classification using deep learning algorithm, and the evaluation of the results.

## THE DATA

Our dataset is composed of light curve images that were constructed by using the eclipsing binary systems observed by the ASAS-SN survey. The light curve data from the survey consist of HJD (Heliocentric Julian Date), magnitude, and flux values with their formal errors for each system. Morphological classes of the systems are also provided on the survey's variable stars database section. We selected Algol, β Lyr, and W UMa type binary stars for our classification.

The times of the primary minimum data in the catalog are used together with given HJDs to calculate the phase values for each system. Then we produced 160x120 pixels images having phase and magnitude axes by using Python code. The total number of light curve images are 16500 (11550 for the training dataset and 4950 for validation) which corresponds to 3850 train and 1650 validation images per class. Figure 2 represents three samples of image data from three morphologies in our training set.

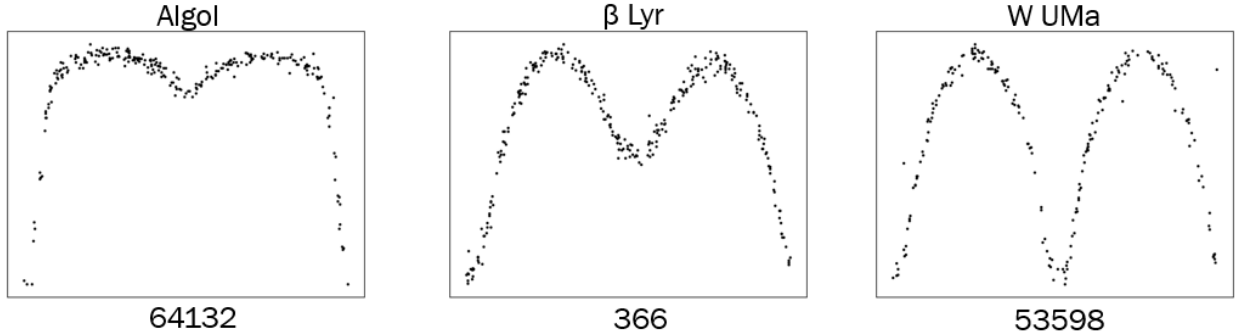

**FIGURE 2**. Sample light curve images from the training data set. Algol type light curve of ID 64132, β Lyr type light curve of ID 366, and W UMa type light curve of ID 53598. These are three of 16500 images produced by using data from the ASAS-SN project.

# THE CODE AND CALCULATIONS

## The Architecture of the Neural Network Model

The code [10] was written in Python [11] programming language on a Jupyter notebook hosted by Google Collaboratory service. Since it aims image classification a neural network was created to achieve the goal. The code begins with importing the necessary modules, packages in which the Scikit-learn machine learning library [12] also takes place. It mainly uses the TensorFlow platform [13] which contains several mathematical libraries to train machine learning models. Keras API [14] (Application Programming Interface) was also used to run TensorFlow in Python language and to be able to apply the code on a GPU (Graphics Processing Unit) since the convolutional networks are computationally expensive in general. Following the importing process, the definitions of the folder paths and image shape are also made.

We define a sequential model that employs the neural network layers linearly. A two-dimensional convolutional neural network which is suitable for image classification was used. The Conv2D layer applies convolution as a specialized linear operation and extracts the features from the input image. During this extraction process, it uses 32 filters with a kernel size of (3, 3), namely $3 \times 3$ matrix ($3 \times 3 \times 3$ in the case of RGB image). Each filter contains a set of weights that determines the output of the convolution layer by scanning the input. The activation layer uses Rectifying Linear Unit (ReLU) which is a ramp function converting negative input values to zero. ReLU provides an output in the form of a linear function and it is an important factor to escape from the complex structure of any nonlinearity. The following convolutional layers whose inputs are the output of the previous layers were added to distinguish edges or sharp structures in images. Max Pooling process is also in charge between layers, therefore, the additional layers deal with the images whose size of the half of the previous images. "EarlyStopping" callback was also used by applying the criteria that stop the program if the validation loss would not decrease in the last ten epochs. The "ModelCheckpoint" callback, on the other hand, saves the model with the maximum validation accuracy up to the present epoch to avoid missing the best model. Fig. 3 shows a schematic representation of the steps in our neural network.

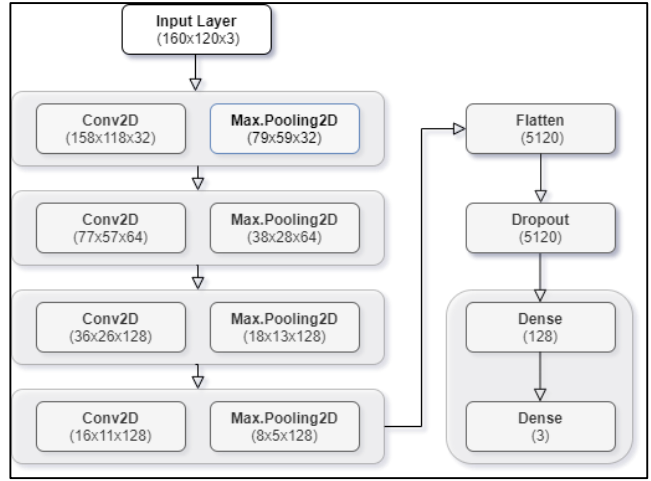

**FIGURE 3.** Schematic representation of the network architecture. The arrows illustrate the workflow through the code. The numbers refer to the sizes of the output images of each layer.

Ninety-six different network architecture was tried for achieving the best result. The trial parameters of these architectures are given in Table 1 with the chosen ones. We obtained the best solution by setting the parameters to the following values: *batch size*:32, *filters*: 32:64:128:128, *L2 regularizer*: 0.001, and the *learning rate*: $10^{-7}$.

**TABLE 1.** The parameters tried on 96 different architecture to achieve the best solution.

| Parameter | Test Values | Chosen values |
|---|---|---|
| Batch size | 16, 32, 64 | 32 |
| Number of 3x3 Filters | 32:32 | 32:64:128:128 |
|  | 32:64 |  |
|  | 32:32:64 |  |
|  | 32:64:64 |  |
|  | 32:64:128 |  |
|  | 32:64:128:128 |  |
|  | 32:64:128:256 |  |
|  | 32:64:128:256:256 |  |
| L2 regularizer | None, 0.001 | 0.001 |
| Learning rate | $10^{-6}$, $10^{-7}$ | $10^{-7}$ |

## Results of the Classification

In a deep learning model, the accuracy can be used in the estimation of correctly predicted data while the loss function refers to a measure of the minimization of the error. Accuracy is basically defined as the ratio of the correct predictions to the total predictions. Our model gives 92% accuracy to an external image that can be clearly seen in Fig. 4 which is a plot of the variation of accuracy and loss values with epoch number. The accuracy increases with the epoch while the cross-entropy loss decreases to about 0.21 for the classification.

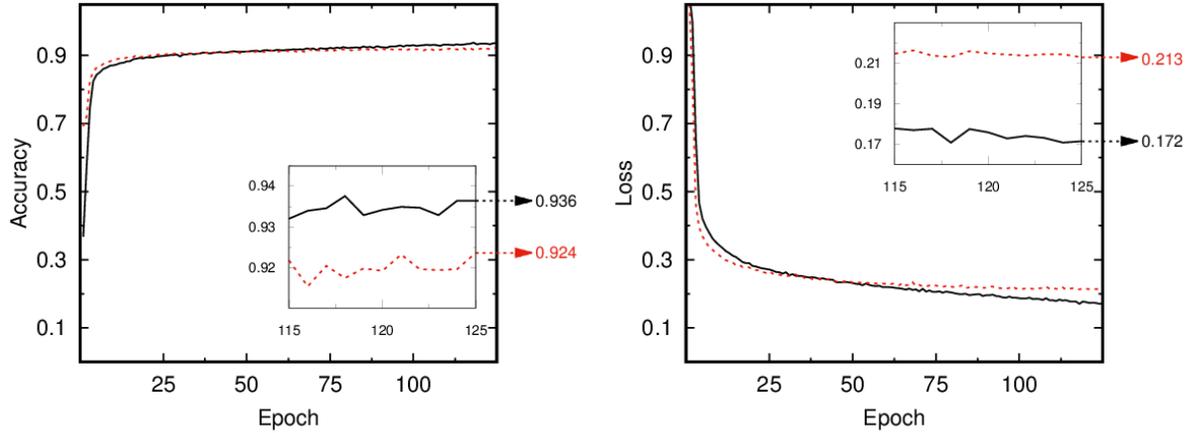

**FIGURE 4.** Variation in accuracy and the loss during the classification. The left and right panels show the variation in accuracy and cross-entropy loss of training (black, solid lines) and validation (red, dotted lines) datasets with the epoch number, respectively. The last ten epoch zoomed inlet for the sake of clear visibility of the variation.

A detailed look at the classification can be made by combining the accuracy values with the other parameters: precision, recall, f1-score which were also calculated for both each class and overall data. The results are listed in Table 2. The weighted average of the precision is achieved 0.92 for the validation dataset which contains 4950 image data from all classes.

**TABLE 2.** The classification report showing the resulting parameters.

|  | Precision | Recall | F1-score | Number of data |
|---|---|---|---|---|
| Algol | 0.93 | 0.93 | 0.93 | 1650 |
| β Lyr | 0.89 | 0.87 | 0.88 | 1650 |
| W UMa | 0.93 | 0.95 | 0.94 | 1650 |
| Weighted average | 0.92 | 0.92 | 0.92 | 4950 |

## CONCLUSION

We presented a deep learning algorithm using convolutional neural networks to classify the morphological classes of ASAS-SN eclipsing binaries' light curves. After obtaining the light curve data from the database, we produce light curve images and grouped them based on their morphological classes. A Python code was written to apply the architecture to 16500 images belonging to three different classes and it is accessible with the data over GitHub [9]. The results show that our architecture can classify morphologies of the light curve images with an accuracy of 92%. The confusion matrix is also extracted (Fig. 5) to represent the correct and incorrect predictions of the algorithm on validation dataset. The numbers of correctly predicted images are 1534, 1432, and 1569 in Algol, β Lyr, and W Uma classes, respectively.

**FIGURE 5.** The confusion matrix showing the predicted and true labels of the image data from the validation dataset.

Since the code is appropriate to adapt further datasets the study is planned to be extended to other databases providing eclipsing binary light curve data. Kepler project [15], CALEB catalog [16], Catalina Survey [17], and OGLE project [18] can be given as an example of these types of observational data. The major difficulty in working on a dataset can be rise from the number of data per class in the training set. At least 1000 images per class are advised in the nonuse of the pre-trained model [19].